%% file: charm_JHEPr1.tex
\documentclass[english,11pt,a4paper]{article}
\pdfoutput=1
\usepackage{jheppub}

\usepackage[T1]{fontenc}
\usepackage[font=small,labelfont=bf]{caption}
\usepackage{enumitem}
\usepackage[force]{feynmp-auto}
\usepackage{hyperref}
\usepackage{cancel}
\usepackage{multirow}
\usepackage{subcaption}
\usepackage{url}
\usepackage{tikz}
\usetikzlibrary{positioning}

\renewcommand{\eqref}[1]{Eq.~(\ref{#1})}

\newcommand{\secref}[1]{Sec.~\ref{sec:#1}}

\newcommand{\Secref}[1]{Section~\ref{sec:#1}}

\newcommand{\figref}[1]{Fig.~\ref{fig:#1}}
\newcommand{\figsref}[2]{Figs.~\ref{fig:#1} and \ref{fig:#2}}

\newcommand{\Figref}[1]{Figure~\ref{fig:#1}}
\newcommand{\tableref}[1]{Table~\ref{tab:#1}}

\newcommand{\beqa}{\begin{align}}
\newcommand{\eeqa}{\end{align}}
\newcommand{\beq}{\begin{equation}}
\newcommand{\eeq}{\end{equation}}

\newcommand{\bpm}{\begin{pmatrix}}
\newcommand{\epm}{\end{pmatrix}}

\newcommand{\parenthesis}[1]{{\left(#1\right) }}
\newcommand{\abs}[1]{{\left|#1\right| }}
\global\long\def\abs#1{\left|#1\right|}

\newcommand\w[1]{_{\mathrm{#1}}}

\newcommand\Order{\mathop{\mathcal{O}}}

\let\gsim\gtrsim

\newcommand\unit[1]{\,\mathrm{#1}}

\newcommand\GeV{\unit{GeV}}
\newcommand\TeV{\unit{TeV}}

\newcommand\ifb{\unit{fb^{-1}}}

\newcommand\mET{\cancel{E}\w T}
\newcommand\pT{p \w T}
\newcommand\HT{H \w T}
\newcommand\meff{m \w {eff}}
\newcommand\mTtwo{m\w{T2}}
\newcommand\mneut[1]{m_{\tilde\chi^0_{#1}}}
\newcommand{\nsq}{N_{{\tilde{q}}}}
\newcommand\dtag{\textrm{2-tag}}

\makeatletter
\def\EE{\@ifnextchar-{\@@EE}{\@EE}}
\def\@EE#1{\ifnum#1=1 \times10 \else \times10^{#1}\fi}
\def\@@EE#1#2{\times10^{-#2}}
\makeatother

\title{Tagging new physics with charm}

\author{S. Iwamoto,}
\author{G. Lee,}
\author{Y. Shadmi,}
\author{and Y. Weiss}
\affiliation{Physics Department, Technion---Israel Institute of Technology,\\ Haifa 32000, Israel}

\emailAdd{sho@physics.technion.ac.il}
\emailAdd{leeg@physics.technion.ac.il}
\emailAdd{yshadmi@physics.technion.ac.il}
\emailAdd{yanivwe@tx.technion.ac.il}

\abstract{%
We propose a new variable, the charm fraction, for collider searches for new physics.
We analyze this variable in the context of searches for simplified supersymmetry models
with squarks, the gluino, and the bino, assuming that only the lightest mass-degenerate squarks
can be produced at the high-luminosity LHC.
The charm fraction complements event counting and kinematic information,
increasing the sensitivity of the searches for models with heavy
gluinos, for which squark production is flavor-blind.
If squarks are discovered at the LHC, this variable can help discriminate between different underlying models.
In particular, with improved charm tagging, the charm fraction can provide information on the gluino mass,
and in some scenarios, on whether this mass is within the reach of a future 100~TeV hadron collider.
}

\begin{document}
\maketitle


\section{Introduction} \label{sec:intro}

Charm tagging at the LHC is a topic of intense study~\cite{ATL-PHYS-PUB-2015-001,CMS-PAS-BTV-16-001},
with future advances expected with the implementation and improvement of machine learning algorithms.
It has recently been added to the menu of supersymmetry searches~\cite{Mahbubani:2012qq,Kalderon:2016wqb,CMS-PAS-SUS-13-009},
along with the much more mature bottom tagging,
leading to improved limits on the masses of charm squarks~\cite{Aad:2015gna}.
In the context of searches for physics Beyond the Standard Model (BSM),
there is a fundamental difference between bottom and charm tagging.
The third generation is expected to play a special role in extensions of the Standard Model (SM): 
light top partners are motivated by naturalness, and at the same time, the masses of top 
and bottom partners may be significantly affected by their large Yukawa couplings.
In contrast, because of the smallness of the relevant Yukawa couplings,
BSM physics may plausibly be first- and second-generation flavor blind.
BSM processes at the LHC may then produce similar amounts of first- and second-generation quarks, 
whereas SM processes are dominated by first-generation quarks.
The fraction of charm quarks in BSM candidate events can thus be a useful
discriminator between new physics and the SM, and between different BSM models.

To examine this, we study the charm fraction in the production of mass-degenerate squark pairs.
We use simplified models, containing only the gluino, some subset of the first- and second-generation squarks, 
and a bino as the lightest supersymmetric particle (LSP).
These models are the ``bread and butter'' of supersymmetry searches,
predicting two or more hard jets and large missing energy ($\mET$).
Generically, the cross section for squark-pair production drops steeply with the gluino mass.
Thus, these searches become more challenging for heavy gluinos.
However, with 8-fold squark degeneracy, squark production becomes more flavor-blind as the gluino mass increases.
For a decoupled gluino, pairs of charm squarks constitute 25\% of the supersymmetry sample.
Measuring this fraction can thus increase the sensitivity of LHC squark searches to scenarios with very heavy gluinos,
which are challenging due to their smaller production cross sections.
As the gluino mass decreases, $t$-channel gluino-exchange diagrams with quarks in the initial state become increasingly important,
and the fraction of charm quarks from pair-produced squarks goes down.

We therefore assume that the gluino is beyond the discovery reach of the high-luminosity LHC (HL-LHC),
and study the charm fraction in squark pair production.
If an excess is observed in jets plus missing energy searches,
significant effort would be required in order to determine the multiplicity,
the SU(3) charge, the mass, and the spin of the particles produced.
In addition, a key question is whether additional particles with masses beyond the LHC reach exist.
The simplified models we study here are characterized by four parameters:
the number of squark flavors produced, the gluino mass, the squark mass, and the bino LSP mass.
The first three determine the BSM production cross section,
while the latter two---and in particular, their difference---determine the
event kinematics and consequently the efficiency of the search.
As is well known, measurements of various kinematic observables
such as the effective mass $m_{\rm{eff}}$ and the stransverse mass $\mTtwo$
can be used to extract some information on the squark and bino masses~\cite{Lester:1999tx}.
The charm fraction, which is a qualitatively different observable, 
can yield new information on the underlying model, and in particular on the gluino mass.
The latter is important input for the planning of future accelerators
like the 100~TeV proton-proton collider, which is anticipated
to be sensitive to gluino masses up to 10--15~TeV~\cite{Cohen:2013xda,Golling:2016gvc}.

While we use squark production as a concrete example,
we believe it is important to approach LHC searches with as few theory biases as possible.
Thus for example, the new states produced could be colored Kaluza-Klein fermions, 
and the ``squarks'' should be thought of merely as new fundamental colored scalars. 
The charm fraction may help determine whether these new produced states are the end of the story. 

This paper is organized as follows. 
In \Secref{framework}, we specify the simplified models we use, review the basics of squark searches, 
describe the Monte Carlo numerical analysis,
and expand on the treatment of charm tagging in our analysis.
In \Secref{charm}, we proceed to study the charm fraction in the models and discuss the results.
We end with some remarks in \Secref{concl}.


\section{General framework: models and overview of searches} \label{sec:framework}

In order to demonstrate the use of the charm fraction, we consider simplified
models, consisting of the first- and second-generation squarks, the gluino, and a bino LSP.
We assume that the squark spectrum is flavor-blind:
squarks of the same gauge quantum numbers are mass degenerate,
while some hierarchies may exist between left-handed and right-handed squarks, 
and/or between up- and down-type squarks.
We imagine a scenario in which the gluino is beyond the reach of the 14~TeV LHC, with mass above 4~TeV~\cite{ATL-PHYS-PUB-2014-010}, 
and assume that only the lightest squarks can be directly produced.
Note that the latter assumption requires only mild hierarchies among the squark masses.

Since the gluino mass affects the production of squark pairs,
the models are characterized by the squark mass $m_{\tilde q}$, the bino mass $\mneut1$, the gluino mass $m_{\tilde g}$, 
and the number of squark flavors produced, which we denote by $\nsq$.
We consider three scenarios:
\begin{itemize}
\item $\nsq=8$ models, in which all the left- and right-handed squarks are degenerate;
\item $\nsq=4$ models, in which only the right-handed squarks can be directly produced
  (with the left-handed squarks beyond LHC reach);
\item $\nsq=2$ models, in which only the right-handed up-type squarks,
  $\tilde u\w{R}$ and $\tilde c\w{R}$, can be directly produced
   (with all remaining squarks beyond LHC reach).
\end{itemize}
The parameters of the different models are summarized in~\tableref{simplifiedModels}.

\begin{table}[t]
 \centering
 \caption{Simplified supersymmetry models used in this paper. \label{tab:simplifiedModels}}
	\begin{tabular}{|c|c|c|c|}\hline
	  $\nsq$ & $m_{\tilde{q}}$ [GeV] & $\mneut1$ [GeV] & $m_{\tilde g}$ [TeV]
	  \\\hline
	  \multirow{4}{*}{8 ($\tilde{u}\w{L,R}, \tilde{c}\w{L,R}, \tilde{d}\w{L,R}, \tilde{s}\w{L,R}$)}
	    & \multirow{2}{*}{1600} & 500 & \multirow{10}{*}{4, 5, 6, 7.5, 10, 13, 450} \\\cline{3-3}
	    &                       & 300 & \\\cline{2-3}
	    & \multirow{2}{*}{1500} & 300 & \\\cline{3-3}
	    &                       &   1 & \\\cline{1-3}
	  \multirow{2}{*}{4 ($\tilde{u}\w R, \tilde{c}\w R, \tilde{d}\w R, \tilde{s}\w R$)}
	    & 1600 & 300 & \\\cline{2-3}
	    & 1400 &   1 & \\\cline{1-3}
	  \multirow{4}{*}{2 ($\tilde{u}\w R, \tilde{c}\w R$)}
	    & 1600 & 300 & \\\cline{2-3}
	    & 1500 & 300 & \\\cline{2-3}
	    & 1500 &   1 & \\\cline{2-3}
	    & 1400 &   1 & \\\hline
	 \end{tabular}
\end{table}

\subsection{Search basics} \label{sec:basics}

The simplified models we consider predict squark-pair production at the LHC, 
yielding events with at least two hard jets, large missing energy, and no electron or muon.
Our analysis below closely follows ATLAS analyses of this topology.
As we focus on the HL-LHC with $\sqrt s=14\TeV$ and an integrated luminosity $\int\mathcal L=3000\ifb$, 
we employ the Meff-2j-3100 signal region (SR) of Ref.~\cite{ATL-PHYS-PUB-2014-010}, which discusses HL-LHC reaches based on this event topology.
This SR selects events with two or more jets, missing energy above 160~GeV, and inclusive effective mass above 3100~GeV.
Some of the model points we will display in the discussion of the charm fraction are already excluded by 13 TeV data. In order to determine whether a model is already excluded, we use the Meff-2j-2000 SR of Ref.~\cite{ATLAS-CONF-2016-078},
which is based on 13~TeV LHC data with $\int\mathcal L=13.3\ifb$.
The full sets of cuts defining both SRs are reviewed in~\tableref{Analysis:bkgEst:SRdefs}.

\begin{table}[t]
 \centering
 \caption{Definitions of our signal regions. SR Meff-2j-2000 is from the ATLAS
   analysis based on $13.3\ifb$ data at the 13~TeV
   LHC~\cite{ATLAS-CONF-2016-078},
   and Meff-2j-3100 is based on the HL-LHC study~\cite{ATL-PHYS-PUB-2014-010}.
   In Meff-2j-2000 (Meff-2j-3100), jets are required to satisfy $\pT>50\GeV$ and
   $|\eta|<2.8$ ($\pT>20\GeV$ and $|\eta|<4.5$), and $\Delta\phi$ cuts are applied
   to all the jets with $\pT>50\GeV$ ($\pT>40\GeV$).
   $\HT$ is the scalar sum of $\pT$ of all the jets,
   and $\meff(\text{incl.})$ is the sum of $\mET$ and $\HT$.
   Events are vetoed if electrons and/or muons with $\pT>10\GeV$ are present.
 }
 \label{tab:Analysis:bkgEst:SRdefs}
 \bgroup
 \def\arraystretch{1.2}%
\begin{tabular}{|cc|c|c|} \hline
  \multicolumn{2}{|c|}{}                      & Meff-2j-2000 & Meff-2j-3100\\\hline
  \multicolumn{2}{|c|}{Number of jets, electrons, muons} & \multicolumn{2}{|c|}{$\ge2$, $=0$, $=0$} \\\hline
  $\mET$ [GeV]                          & $>$ & 250      & 160    \\\hline
  \mbox{\quad}$\pT(j_1), \pT(j_2)$ [GeV] \mbox{\quad}           & $>$ & 250, 250 & 160, 60 \\\hline
  $\abs{\eta(j_1, j_2)}$                & $<$ & 1.2      & ---    \\\hline
  $\Delta\phi(j_{1,2,(3)},\mET)\w{min}$ & $>$ & 0.8      & 0.4    \\\hline
  $\Delta\phi(j_{i>3},\mET)\w{min}$     & $>$ & 0.4      & 0.2    \\\hline
  $\mET/\sqrt{\HT}$ [GeV$^{1/2}$]       & $>$ & 20       & 15     \\\hline
  $\meff(\text{incl.})$ [GeV]           & $>$ & 2000     & 3100   \\\hline
\end{tabular}
\egroup
\end{table}

\subsection{SM backgrounds and charm production}

The main SM background for the two-jets plus missing energy search
is $Z+\text{jets}$ with the $Z$ boson decaying into neutrinos (see \figref{bkgZjGluonSplitting} and \figref{bkgZjqgprod}).
The next source of background is $W+\text{jets}$ production, which we return to below. 
Dibosons and $t\bar{t}$ production give smaller contributions (see, e.g., Ref.~\cite{ATLAS-CONF-2016-078}).

The dominant $W$-background is from $W$ decays to tau plus neutrino,
with the tau decaying hadronically, predominantly to light jets.
These processes receive large contributions from diagrams like~\figref{bkgZjqgprod},
with the $Z$ replaced by a $W$, leading to a $\tau$, neutrino and two jets in the final state,
as well as from diagrams like~\figref{bkgWjTau}, which lead to a $\tau$, neutrino and one jet.
Another type of background from $W$ production is processes in which
the $W$ decays into a light lepton (electron or muon) plus neutrinos,
and the lepton is lost in the reconstruction (see, e.g.,~\figref{bkgWjLostLepton}).

In the invisible-$Z$ background, the leading source of charm quarks is QCD production of $c\bar c$ pairs as shown in \figref{bkgZjGluonSplitting}.
Another important source of charm quarks in the SM background comes from higher-order, but log-enhanced, processes, 
and in particular ``gluon splitting'' into a $c\bar{c}$ pair (see, e.g., Ref.~\cite{Banfi:2007gu}). 
While this is not included in our leading order (LO) simulation of the hard processes, some component of charm pairs from gluon splitting is generated by \texttt{Pythia}.
Gluon-charm initial states in \figref{bkgZjqgprod} and gluon-strange initial states in \figsref{bkgWjTau}{bkgWjLostLepton} give small contributions since they are PDF-suppressed. 
In the latter two figures, processes with an initial-state down-quark are CKM-suppressed.

\begin{figure}[t]
  \centering
  \begin{subfigure}[t]{0.35\textwidth}
    \centering
    \input{feyn_bkg_zj_gluon.tex}
    \caption{$Z+ q \bar{q}$ production \label{fig:bkgZjGluonSplitting}}
 \end{subfigure}
 \begin{subfigure}[t]{0.35\textwidth}
  	\centering
  	\input{feyn_bkg_zj_quark.tex}
  	\caption{$Z+ q g$ production \label{fig:bkgZjqgprod}}
 \end{subfigure}
 \caption{\label{fig:zPlusJets}Diagrams contributing to the $Z+\text{jets}$ background.}
\end{figure}
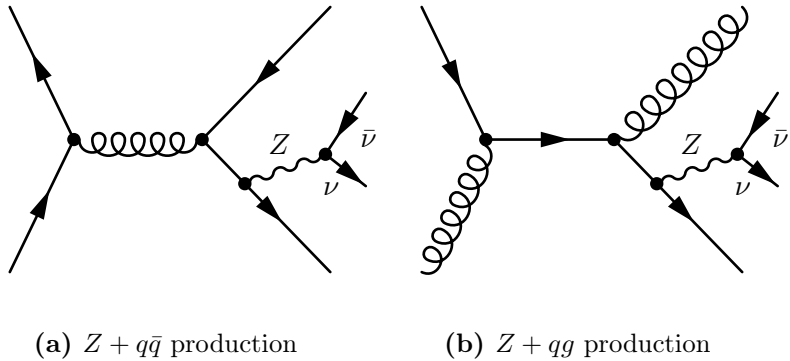

\begin{figure}[t]
  \centering
  \begin{subfigure}[t]{0.35\textwidth}
    \centering
  	\input{feyn_bkg_wj_tau.tex}
   	\caption{Hadronic tau decay \label{fig:bkgWjTau}}  
  \end{subfigure}
  \begin{subfigure}[t]{0.35\textwidth}
    \centering
   \input{feyn_bkg_wj_lostlepton.tex}
    \caption{Lost lepton ($\ell = e, \mu$) \label{fig:bkgWjLostLepton}}
  \end{subfigure}
  \caption{\label{fig:wPlusJets}Diagrams contributing to the $W+\text{jets}$ background.
    $\cancel{\ell}$ denotes an electron or muon that is lost in the reconstruction.}
\end{figure}
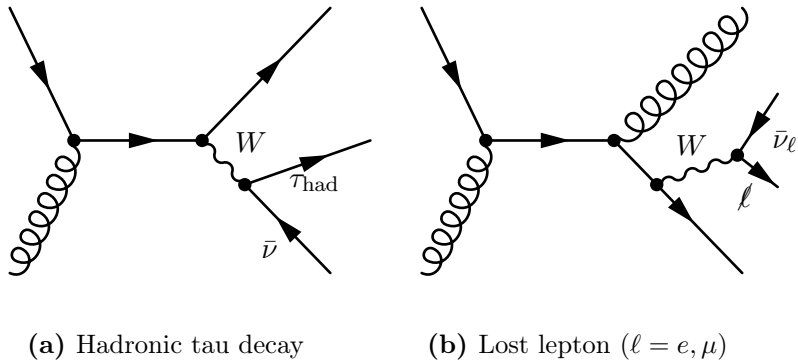

\subsection{Monte Carlo simulation} \label{sec:mc}

Signal samples are generated by \texttt{MadGraph\_aMC@NLO~5}~\cite{Alwall:2014hca} at LO,
with the PDF set \texttt{NNPDF2.3QED} at LO with $\alpha_s=0.13$~\cite{Ball:2013hta}.
The baseline selections described in~\tableref{Analysis:MG5Cuts} are applied based on the missing transverse energy and jet $\pT$.
Parton showering and hadronization are performed by \texttt{Pythia~6.4}~\cite{Sjostrand:2006za}.
Tau decays are simulated by \texttt{TAUOLA}~\cite{Jadach:1993hs}.
For detector simulation, \texttt{Delphes~3.3.0}~\cite{deFavereau:2013fsa} is utilized with the default detector card, 
where the parameter of anti-$k\w T$ algorithm~\cite{Cacciari:2008gp,Cacciari:2011ma} for jet clustering is replaced by $R=0.4$ to match the ATLAS studies.
Pile-up effects are not considered.
These signal samples are rescaled by next-to-leading-order (NLO) $K$-factors, 
which are calculated by \texttt{Prospino~2}~\cite{Beenakker:1996ch}%
\footnote{\texttt{Prospino} does not handle non-degenerate squarks from the first two generations. 
However, the NLO correction is dominated by QCD contributions (light quarks and gluons)~\cite{GoncalvesNetto:2012yt}. 
Therefore, additional heavy squarks only contribute at next-to-next-to-leading order and 
the \texttt{Prospino} $K$-factors are a good approximation even in this case.
As previously mentioned, we only require mild hierarchies between the masses of the lightest and other squarks, so we will ignore leading-log corrections.}.
We then apply the selection cuts for SR Meff-2j-2000 and, using the $K$-factors for 13 TeV collisions, and compare to the upper
bound obtained by the ATLAS analysis~\cite{ATLAS-CONF-2016-078} to determine whether the model point is excluded.

To obtain the charm fraction at the HL-LHC, background events are also generated by the same procedure.
Our selection cuts, especially the high $\meff$ cut of 3100~GeV, suppress the $W+\text{jets}$ background
so that it is about a third of the $Z+\text{jets}$ background.
At the same time, the $Z+\text{jets}$ background is easier to calculate compared to the $W+\text{jets}$ background, 
since the latter comprises different components  (e.g., $j\tau\nu$, $jj\tau\nu$).
We simulate the $Z+\text{jets}$ background and the different components of the $W+\text{jets}$ background at leading order.
We find that the fractions of (truth-level) charm quarks in each of these are similar.
We therefore obtain the total number of background events by
reweighting the $Z+\text{jets}$ sample (with the baseline selection in \tableref{Analysis:MG5Cuts}) 
to match the number of events from $Z+\text{jets}$ and $W+\text{jets}$ processes 
in the ATLAS analysis~\cite{ATL-PHYS-PUB-2014-010} (Figure 8b),
and approximate the fraction of charm quark events (whether these contain a single charm quark or a pair of charm quarks) 
in this sample by its value in our simulated  $Z+\text{jets}$ sample.

Events are selected by the cuts of the Meff-2j-3100 SR.
For each event in the SR, the two jets with leading $p\w{T}$ are considered in the calculation of the charm fraction.
Initially, \texttt{Delphes 3.3.0} is utilized to determine the ``truth-level'' jet flavor.
In this algorithm, a jet is considered bottom-flavored, or a ``truth-level bottom jet'', 
if one or more bottom quarks ($b$ or $\bar b$) exist in the jet cone as a \texttt{Pythia}-level parton.
If no bottom partons are found but charm partons exist, the jet is labeled as a ``truth-level charm jet''.
Otherwise, the jet is treated as a light jet at the truth level.
The detector-level charm tagging is performed based on this ``truth-level'' information as we describe below.

\begin{table}[t]
 \centering
 \caption{Baseline event selection we use in the simulation. Variables are calculated at the \texttt{MG5\_aMC} level. The missing transverse momentum $\cancel{p}\w T$ is defined by the sum of neutrino momenta.}
\begin{tabular}{|cc|c|c|c|c|}\hline
  \multirow{2}{*}{}            &     & Meff-2j-2000 & \multicolumn{2}{c|}{Meff-2j-3100} \\\cline{3-5}
                               &     & Signals      & $Z+\text{jets}$ & Signals \\\hline
  $|\cancel{p}\w T|$   [GeV]   & $>$ & ---          & 150             & ---     \\\hline
  Leading jet $p\w T$    [GeV] & $>$ & 150          & 500             & 150     \\\hline
  Subleading jet $p\w T$ [GeV] & $>$ & ---          & 60              & ---     \\\hline
\end{tabular}
\label{tab:Analysis:MG5Cuts}
\end{table}

\subsection{Charm tagging} \label{sec:tagging}

The main limiting factor in measurements of the charm fraction is
the charm tagging capabilities and in particular the fake rates.
Current analyses at the LHC experiments utilize charm-tagging algorithms based on the working points
$(\epsilon_{c},\epsilon_{b}, \epsilon_{l}) =(0.19, 0.2, 0.005)$~\cite{ATL-PHYS-PUB-2015-001} or $(0.2, 0.24, 0.02)$~\cite{CMS-PAS-BTV-16-001}.
Here $\epsilon_c$ is the tagging efficiency of charm quarks,
while $\epsilon_{b}$ and $\epsilon_{l}$ are the mistag rates for bottom and light jets, respectively.
These taggers are primarily trained on $t\bar{t}$ samples and thus the maximal
jet transverse momentum does not exceed 300~GeV~\cite{ATL-PHYS-PUB-2014-010}.
In contrast, the average $\pT$ for the simplified models presented here
is $\sim 500\GeV$, so charm-tagging would be more challenging.
Still, charm tagging algorithms will likely undergo significant improvement by the end of the HL-LHC
program. Thus, we consider two optimistic scenarios with efficiencies
$\epsilon_c=0.5$ or 0.3 and mistag rates $\epsilon_{b}=0.2$ and $\epsilon_{l}=0.005$
(see also Ref.~\cite{Perez:2015lra}).
Since we cannot reliably estimate the $p\w T$ and $\eta$ dependence of the various efficiencies,
we take them to be constant over the entire ranges.

For a given set of tagging parameters, $(\epsilon_{c},\epsilon_{b}, \epsilon_{l})$, 
we can divide the sample of events passing the cuts as follows.
In each event, we examine the truth-level flavor of each of the two hardest jets.
A truth-level charm jet is ``tagged'' as a charm jet with probability
$\epsilon_{c}$. Similarly, each truth-level $l$ ($b$) jet is ``tagged''
as a charm jet with probability $\epsilon_{l}$ ($\epsilon_{b}$).
We denote the number of these ``charm-tagged'' jets by $N_c$, and the total number of events in the sample by $N\w{ev}$.
For high-efficiency, high-purity taggers, it would also be useful to
separately consider events in which the two hardest jets are both tagged as charm jets. 
We denote the number of these double-tagged events by $N^{\dtag}$.

One caveat of our simulation is the fact that the \texttt{Delphes} algorithm that we utilize treats 
$c\bar c$ pairs originating from gluons or from taus as charm jets,
even when the pair is clustered into a single jet. 
We return to this issue in the discussion of the results.


\section{The charm fraction} \label{sec:charm}

Squark pairs would be produced at the LHC either through flavor-democratic
processes (see~\figref{democratic}),
or through gluino-mediated processes (see~\figref{unDemocratic}),
which are sensitive to the proton PDF's and are thus flavor-dependent.
As the gluino mass is increased, the latter processes become less significant.

\begin{figure}[t]
 \centering
 \input{feyn_democratic_production.tex}
 \caption{\label{fig:democratic}Diagrams contributing to flavor-democratic squark-pair production.}
\end{figure}
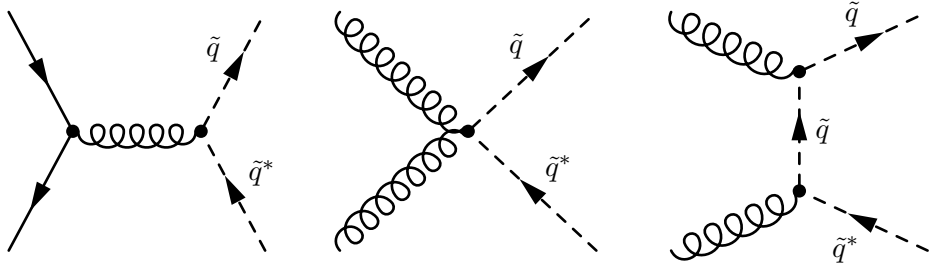
\begin{figure}[t]
 \centering
 \input{feyn_undemocratic_production.tex}
 \caption{\label{fig:unDemocratic}A diagram contributing to flavor-undemocratic squark-pair production.}
\end{figure}
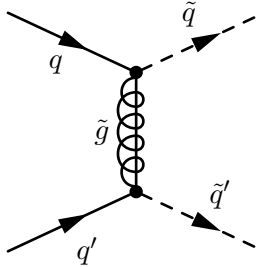

We define the \textit{charm fraction} $F_c$ as the ratio
\begin{equation}
 F_{c} \equiv \frac{N_c}{2N\w{ev}} \,,
\end{equation}
where $N\w{ev}$, $N_c$ were defined in \secref{tagging}.
For events coming from squark pair production, we expect this fraction to increase as the gluino mass increases.
This behavior is exhibited in \figref{truthLevelFcVsMgluino}, 
where we plot the charm fraction for a model with $\nsq=8$ squarks with $m_{\tilde{q}} = 1.5\TeV$ and a massless bino,
assuming an ideal tagger $(\epsilon_c=100\%, \epsilon_b=0,\epsilon_l=0)$.
Error bars are the Monte Carlo uncertainties.
The fraction rises from 0.10 to 0.16 as the gluino mass varies from 4~TeV to 13~TeV,
and asymptotes to 0.25 for a decoupled gluino.
Repeating this for the SM background yields a charm fraction of 0.09.
The hollow points are already excluded by the Meff-2j-2000 SR of~\cite{ATLAS-CONF-2016-078} (see \secref{mc} for details).

\begin{figure}[t]
 \centering
 \begin{subfigure}{0.7\textwidth}
 \centering
 \includegraphics[width=\textwidth]{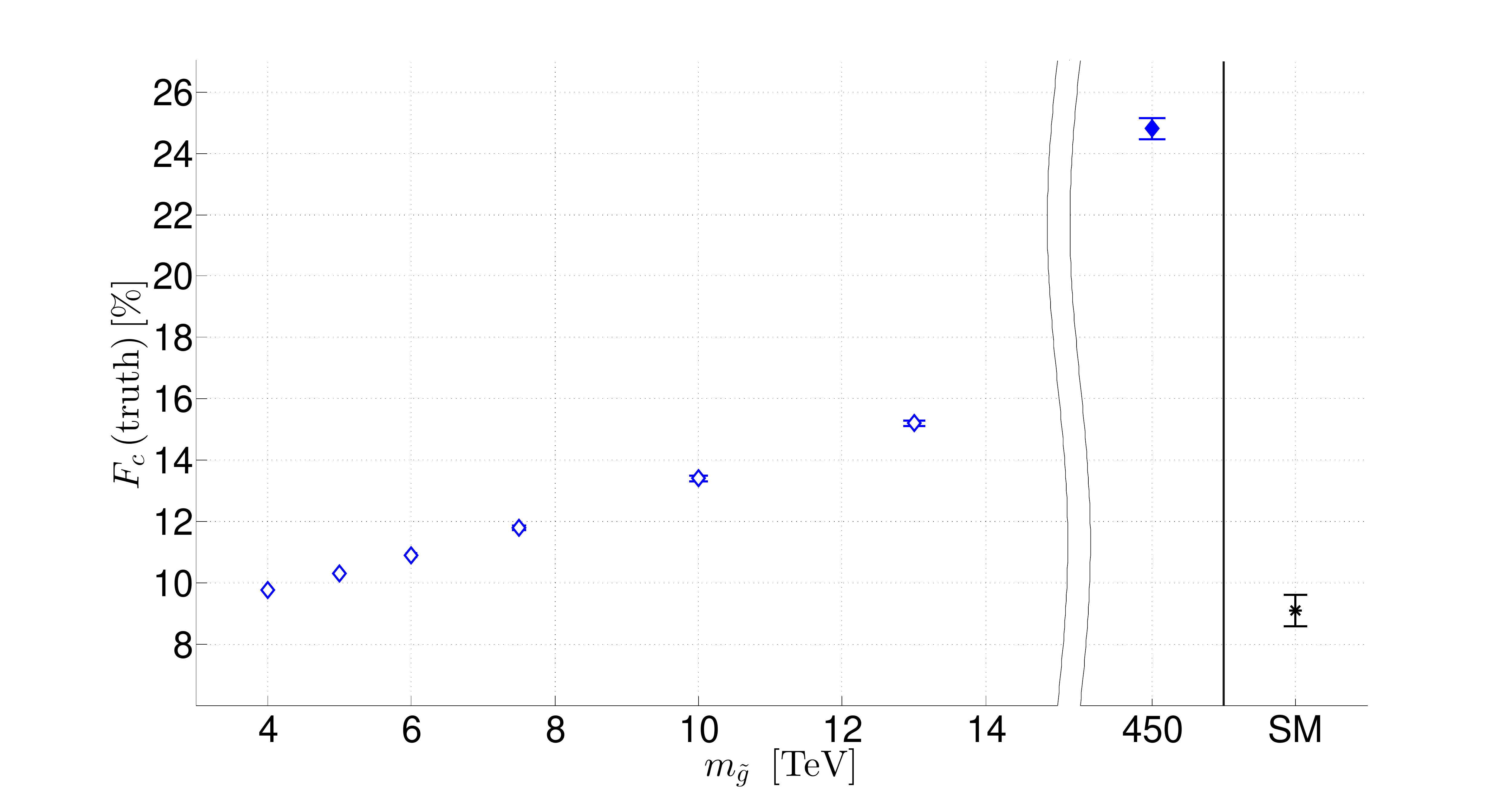}
 \caption{Truth-level charm fraction for jets. \label{fig:truthLevelFcVsMgluino}}
 \end{subfigure}
 
 \begin{subfigure}{0.7\textwidth}
 \centering
 \includegraphics[width=\textwidth]{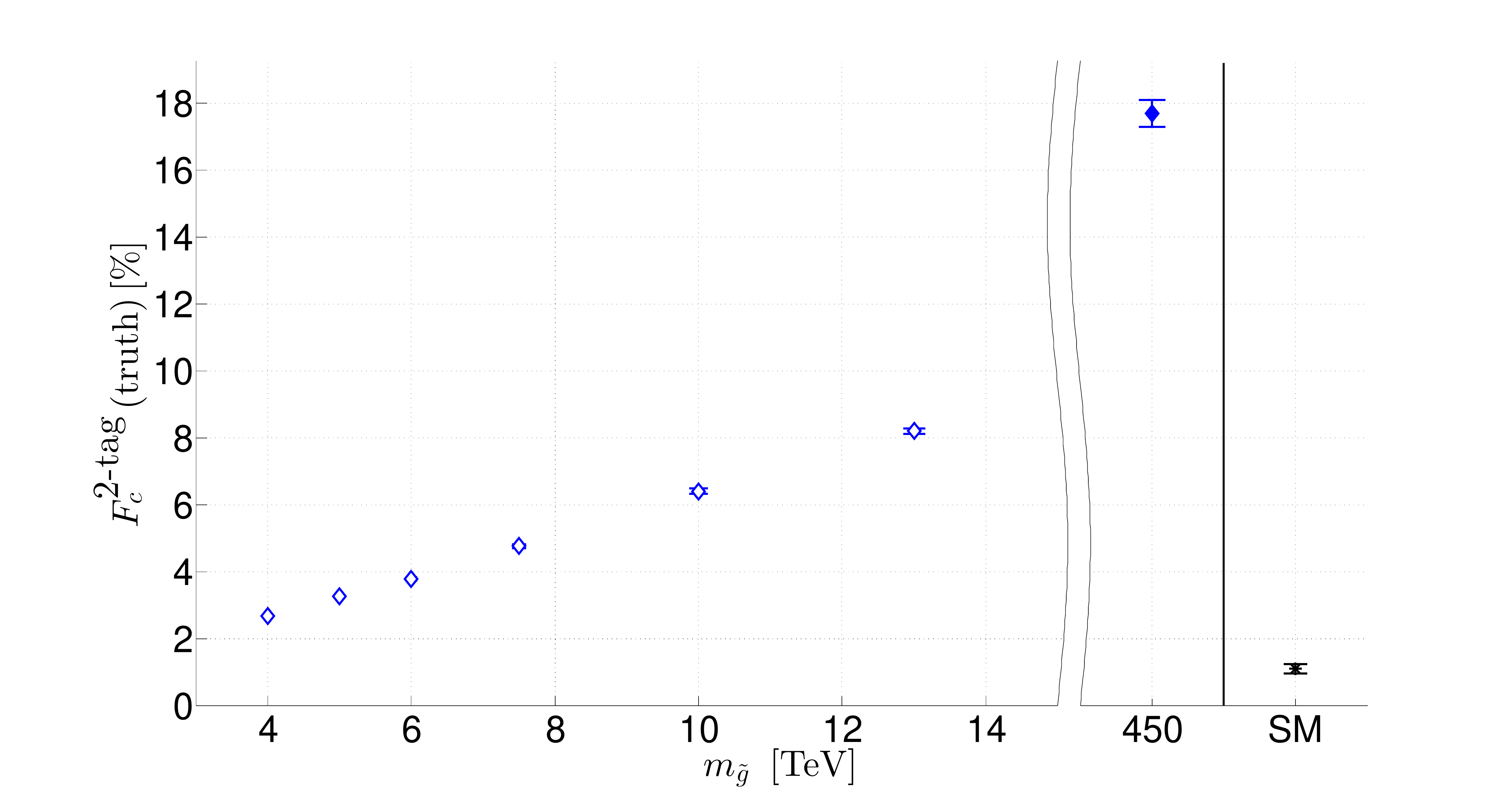}
 \caption{Truth-level charm fraction for double-tagged events. \label{fig:truthLevelRcVsMgluino}}
 \end{subfigure}
  \caption{The truth-level charm fraction for jets (upper) and for double-tagged events (lower) of the signal-only samples as a function of the gluino mass for models with $\nsq=8$, $m_{\tilde q} =1.5\TeV$, and a massless bino, and that of the SM background sample.
  Hollow points are excluded by the ATLAS analysis~\cite{ATLAS-CONF-2016-078}.
  Error bars are the Monte Carlo uncertainties.}
\end{figure}

In \figref{truthLevelRcVsMgluino}, we show the fraction of double-tagged events,
\beq
F_c^{\dtag}\equiv \frac{N^{\textrm{2-tag}}}{N\w{ev}} \,.
\eeq
The value for the decoupled gluino is 0.18, which is smaller than the naive expectation of 0.25 
because jets from QCD radiation can be harder than charm jets from charm squarks.
The SM background is reduced by a larger relative margin as the number of double charm events is smaller for the SM.

\begin{figure}[t]
 \centering
 \includegraphics[width=\textwidth]{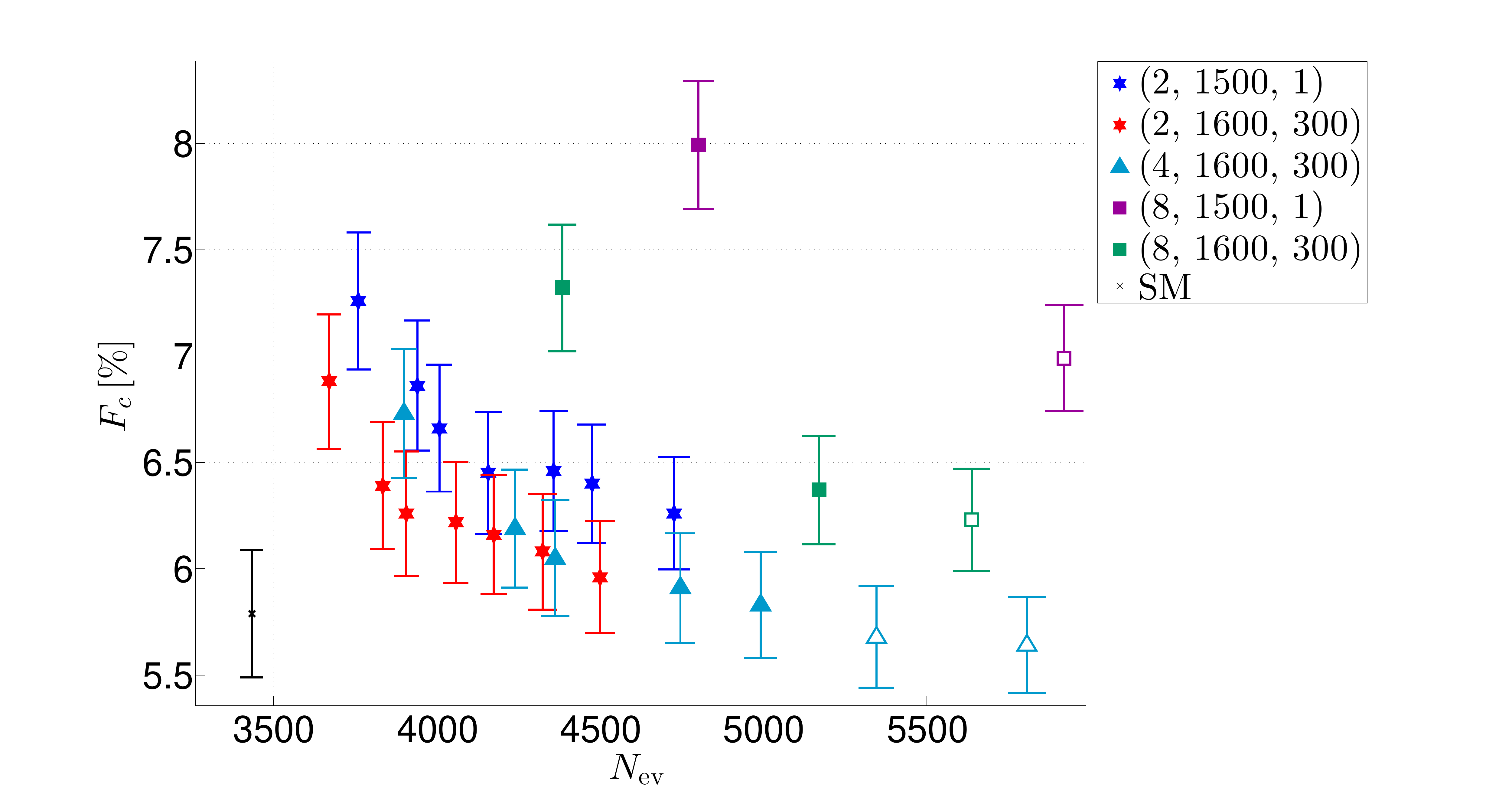}
 \caption{\label{fig:scenario_1_charm_fraction_MT2_1500_Eff_c_50} The number of events $N\w{ev}$ and the charm fraction $F_c$ in the SM+supersymmetry samples, as well as the SM-only sample, in the Meff-2j-3100 SR, expected at the HL-LHC with $\sqrt s=14\TeV$ and $\int\mathcal L=3000\ifb$. The bars are the statistical uncertainty on $F_c$.
   Hollow points are excluded by the ATLAS analysis~\cite{ATLAS-CONF-2016-078}. The tagging efficiencies are $(\epsilon_c, \epsilon_b, \epsilon_l) = (0.5, 0.2, 0.005)$. The theoretical $\mTtwo$ endpoints for these models are in the range $1500\pm50 \GeV$. The numbers in the legend correspond to $(\nsq, m_{\tilde q}/\text{GeV}, \mneut{1}/\text{GeV})$.}
\end{figure}

The results of~\figsref{truthLevelFcVsMgluino}{truthLevelRcVsMgluino} are based on truth-level parton flavor; 
however, realistically, we must consider charm-tagged jets.
In~\figref{scenario_1_charm_fraction_MT2_1500_Eff_c_50}, 
we show the results for various model points in the $N_{\w{ev}}$--$F_c$ plane,
assuming tagging efficiencies of $(\epsilon_{c},\epsilon_{b}, \epsilon_{l}) =(0.5, 0.2, 0.005)$.
We focus on a narrow range of squark masses, for which the models may potentially be probed by the HL-LHC, 
and which are not yet excluded for at least some gluino masses. 
We then vary the number of squarks produced, over $\nsq=2,4,8$, and consider both heavy (300~GeV) and massless binos.
The squark and bino masses are chosen such that all the models yield similar kinematics,
and cannot be distinguished based on $\mTtwo$ \cite{Lester:1999tx}.
Each shape-color combination maps to a particular choice of ($\nsq$, $m_{\tilde q}$, $\mneut1$): 
the shape of the central value marker indicates $\nsq$, 
and the color designates pairs of squark and bino masses ($m_{\tilde q}$, $\mneut1$). 
Points with the same shape and color correspond to the different gluino masses of~\tableref{simplifiedModels}:
in a sequence increasing in $N\w{ev}$, the values of $m_{\tilde g}$ decrease, beginning with $m_{\tilde g} = 450$~TeV.
Points with hollow central values are already excluded by the Meff-2j-2000 analysis of~\cite{ATLAS-CONF-2016-078}.
Only statistical uncertainties on $F_c$ (assuming $3000\ifb$ of integrated luminosity) are shown.

For the largest gluino masses, discovery based on $N\w{ev}$ alone would be challenging.
In~\figref{scenario_1_charm_fraction_MT2_1500_Eff_c_50}, we have not displayed horizontal error bars on $N\w{ev}$
since we cannot reliably estimate the dominant systematic uncertainties.
Still, using current LHC analyses as a guide, it is reasonable to expect
the systematic uncertainty on $N\w{ev}$ to be around 10\%~\cite{ATL-PHYS-PUB-2014-010}.
The SM prediction is then $N\w{ev}^{\mathrm{SM}}=3433\pm 59\w{stat}\pm 343\w{syst} = 3433\pm 348$.
Roughly, most of the model points of~\figref{scenario_1_charm_fraction_MT2_1500_Eff_c_50}
lead to excesses in $N\w{ev}$ below 3$\sigma$ (corresponding to $N\w{ev}\lesssim{4500}$).
Furthermore, for fixed values of $(\nsq, m_{\tilde q}/\text{GeV}, \mneut{1}/\text{GeV})$, 
only a limited range of gluino masses remains for which a 5$\sigma$ discovery, 
requiring $N\w{ev}\gsim5200$, would be possible.
Recall that hollow points denote models which are excluded by Ref.~\cite{ATLAS-CONF-2016-078}.

On the other hand, for large gluino masses, the charm content of supersymmetry events is large,
so charm tagging can be used to increase the sensitivity to these models.
Since it is down by the fraction of charm squarks produced and the charm tagging efficiency, 
the number of charm-tagged events is prone to larger statistical uncertainties compared to $N\w{ev}$;
however, many systematic uncertainties cancel out in this ratio, including the uncertainty on the jet energy scale,
which affects the determination of both the missing energy and $\meff$.
We expect the dominant remaining sources of uncertainties to be the charm tagging efficiencies and the PDF's.
Note that the latter do not completely cancel in the ratio,
as the charm fraction is sensitive to the relative sizes of the PDF's of
the valence quarks, gluons, and sea quarks%
\footnote{The PDF uncertainties are expected to shrink by the end of the HL-LHC program (see, e.g., Ref.~\cite{dEnterria:2017}).}.
As seen in~\figref{scenario_1_charm_fraction_MT2_1500_Eff_c_50},
for models with a heavy gluino, the charm fraction displays the largest deviation from the SM background.

Let us assume a 10\% uncertainty on $F_c$ to get a rough estimate of the discriminating power of this variable.
The SM then predicts $F_c=(5.8\pm0.3\w{stat}\pm0.6\w{syst})\%=(5.8\pm0.6)\%$.
The deviation of $F_c$ from the SM prediction is then at the level of 1.4--3.2$\sigma$ for decoupled gluino models,
and when combined with  $N\w{ev}$, may allow for discovering these models.
Obviously, the $F_c$ values shown in~\figref{scenario_1_charm_fraction_MT2_1500_Eff_c_50} 
are weighted averages of the SM and supersymmetry samples,
and are particularly skewed towards the smaller SM value when the squark production cross section is small.
Thus for example,
for the $\nsq=2$ models with $m_{\tilde q}=1.5$~TeV,
a vanishing bino mass, and gluino masses of 6~TeV and above,
the excesses in $N\w{ev}$ vary between $0.5$--$2.7\sigma$,
while the deviations of the charm fraction from the SM prediction vary between
$2.3\sigma$ (for the decoupled gluino) and $1.0\sigma$ (for the 6~TeV gluino).
The combination of these two variables increases the sensitivity for these challenging scenarios.

If an excess in $N\w{ev}$ is observed, attention will be focused on the properties of the new particles produced and on whether additional new particles exist.
As explained above, the model points shown here are chosen such that the end points of their various $\mTtwo$ distributions lie in the range $1500\pm50$~GeV%
\footnote{The central value of 1500 GeV roughly corresponds to the limit for discovery of squarks at the HL-LHC 
in simplified models with a decoupled gluino, $\nsq = 8$, and a massless bino LSP~\cite{ATL-PHYS-PUB-2014-010}.}. 
Thus, it will be difficult to distinguish between them based on their missing energy signatures. 
The charm fraction can clearly break the degeneracy between different underlying models. 
While a definitive statement cannot be made given that we cannot reliably estimate the systematic uncertainties,
the results suggest that models with gluino masses around or below 10~TeV
can be discriminated from decoupled gluino scenarios. 
Thus for example, $F_c$ can easily discriminate between the $(8,1500,1)$ model with a decoupled gluino, which gives $N\w{ev}\simeq4800$ and $F_c=(7.9\pm0.3\w{stat})\%$, and the other models which give a similar value
of $N\w{ev}$, but with gluino masses between 4--7.5~TeV and $F_c=(5.9\text{--}6.3)\%$; assuming a $10\%$ systematic uncertainty on $F_c$, these are $2\text{--}3\sigma$ away from each other.

\begin{figure}[t]
	\centering
	\includegraphics[width=\textwidth]{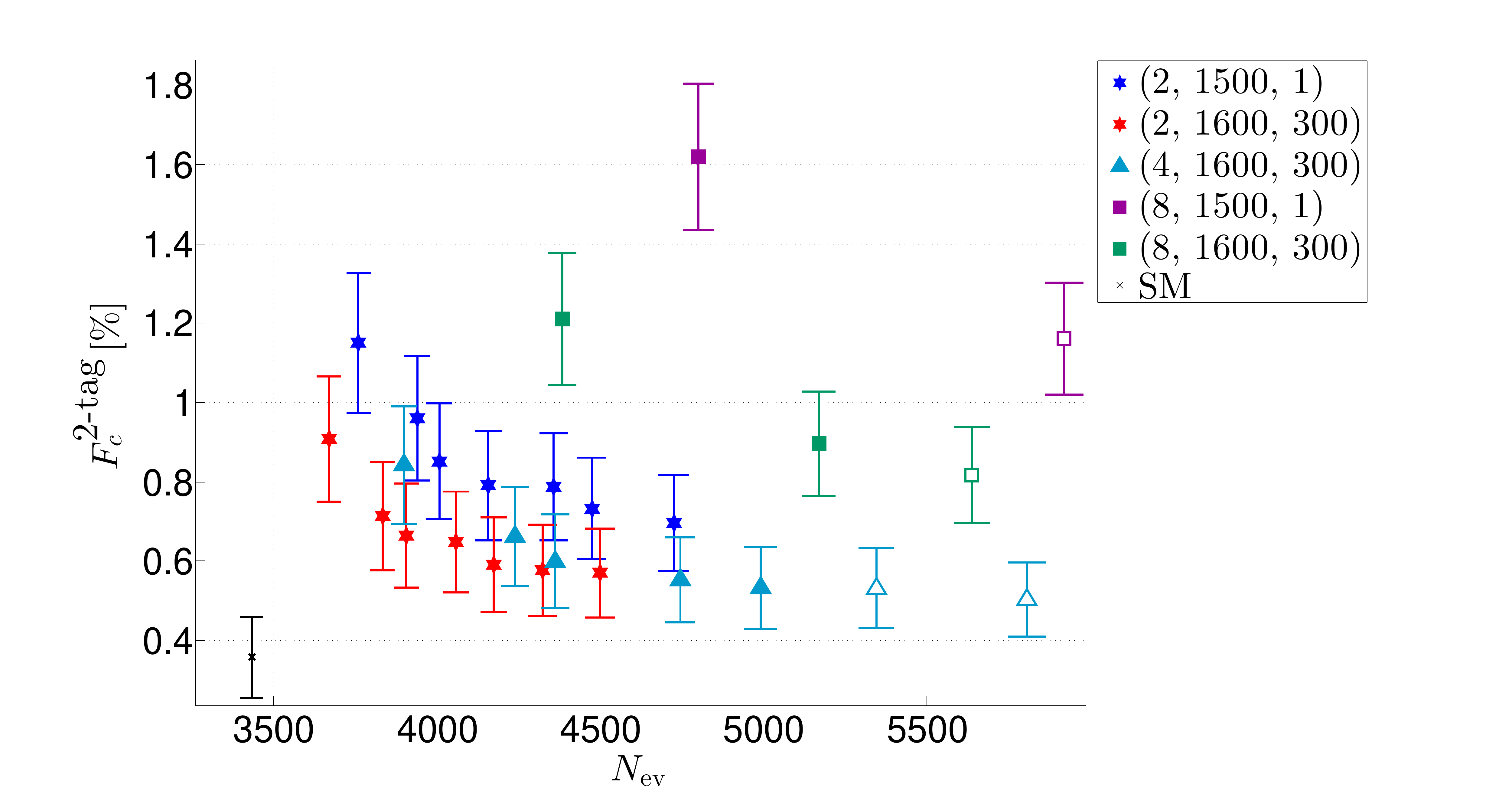}
	\caption{\label{fig:scenario_1_charm_fraction_MT2_1500_Eff_c_50_Rc} Same as \figref{scenario_1_charm_fraction_MT2_1500_Eff_c_50}, but with the fraction of double-tagged events $F^{\dtag}_c$ as the vertical axis.}
\end{figure}

With a high efficiency to tag charm jets, it is sensible to also consider the
fraction of double-charm-tagged events, $F_c^{\dtag}$,
which we plot for the same set of models in \figref{scenario_1_charm_fraction_MT2_1500_Eff_c_50_Rc}.
Compared to $F_c$, $F_c^{\dtag}$ suffers from QCD radiation effects and larger statistical uncertainties 
due to a further reduction by approximately $\epsilon_c$.
If the systematic uncertainty in the charm fraction is dominated by uncertainties on tagging efficiencies,
$F_c^{\dtag}$ will also be subject to a systematic uncertainty approximately twice that of $F_c$.
However, because the SM prediction for $F_c^{\dtag}$ is small,
a deviation from the SM value will be more significant.
For example, assuming a 20\% systematic uncertainty on $F_c^{\dtag}$, the decoupled gluino scenarios with $\nsq=2$ will have $F_c^\dtag=(1.2\pm0.3)\%$ and $(0.96\pm0.25)\%$, which are 4--5$\sigma$ away from the SM expectation.
As for discriminating between different supersymmetry models, comparing \figsref{scenario_1_charm_fraction_MT2_1500_Eff_c_50}{scenario_1_charm_fraction_MT2_1500_Eff_c_50_Rc}, we find that $F_c$ and $F_c^\dtag$ have approximately the same analyzing power.
Considering the aforementioned model points with $N\w{ev}\sim4800$,
the $(8, 1500, 1)$ model with a decoupled gluino gives
$F_c^\dtag=(1.58\pm0.18\w{stat}\pm0.32\w{syst})\%=(1.58\pm0.36)\%$,
while the other models give $(0.60\pm0.16)\%$ and $(0.72\pm0.19)\%$,
which are $2\text{--}3\sigma$ away from each other.

\begin{figure}[p]
 \centering
 \includegraphics[width=\textwidth]{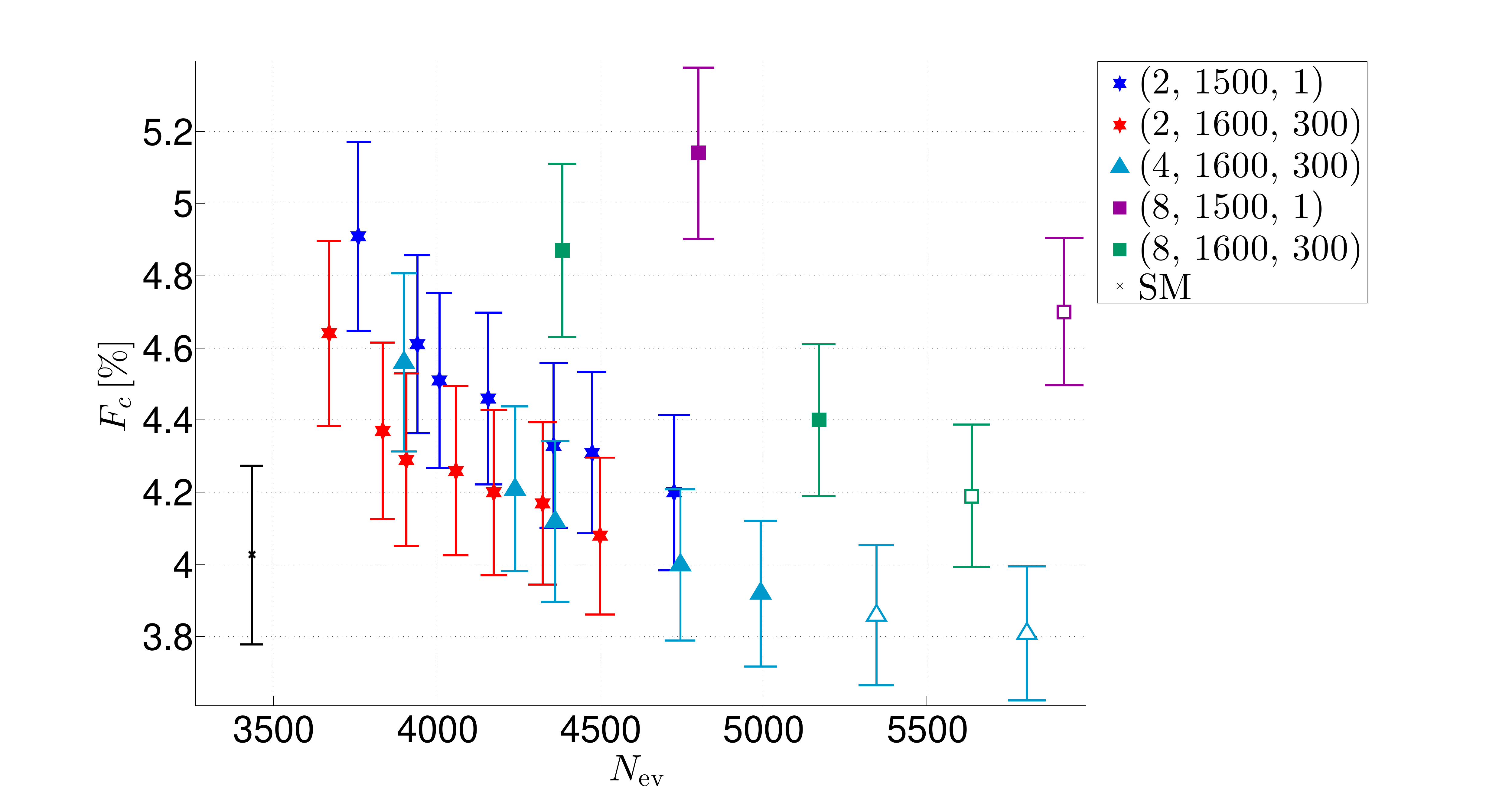}
 \caption{\label{fig:scenario_1_charm_fraction_MT2_1500_Eff_c_30}  Same as \figref{scenario_1_charm_fraction_MT2_1500_Eff_c_50}, but with tagging efficiencies $\parenthesis{\epsilon_{c}, \epsilon_{b}, \epsilon_{l}} = (0.3,0.2,0.005)$.}
\end{figure}
\begin{figure}[p]
	\centering
	\includegraphics[width=\textwidth]{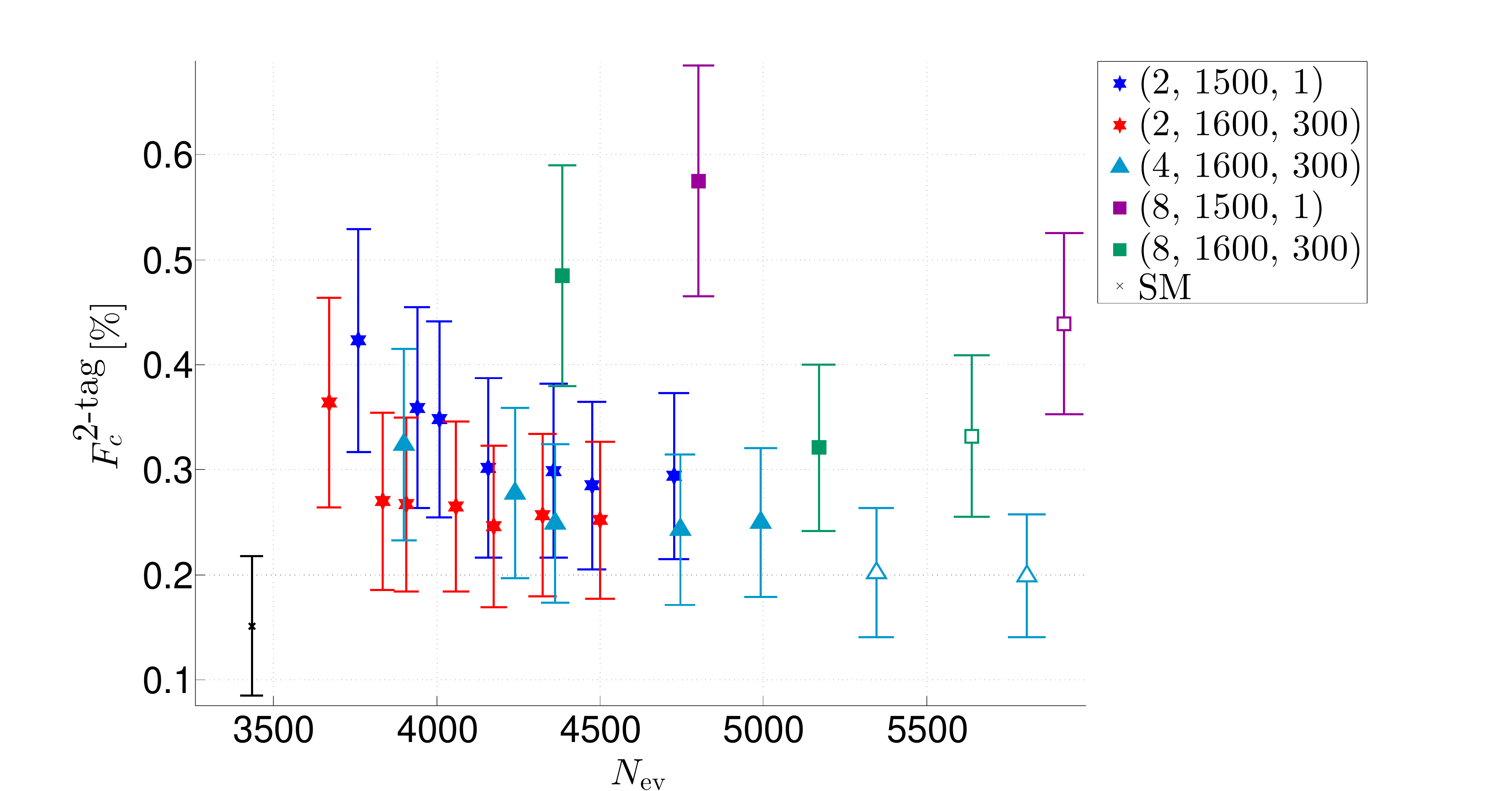}
	\caption{\label{fig:scenario_1_charm_fraction_MT2_1500_Eff_c_30_Rc}  Same as \figref{scenario_1_charm_fraction_MT2_1500_Eff_c_30}, but for double tagged events. }
\end{figure}

\Figref{scenario_1_charm_fraction_MT2_1500_Eff_c_50} assumes charm tagging efficiencies of
$\left(\epsilon_{c}, \epsilon_{b}, \epsilon_{l} \right) = \left(0.5, 0.2, 0.005 \right)$. 
This is far better than those currently published \cite{ATL-PHYS-PUB-2015-001,CMS-PAS-BTV-16-001}. 
Furthermore, these numbers are expected to deteriorate as the jet $\pT$ increases;
however, since we do not know the high-$\pT$ and $\eta$ dependence of the tagger at the HL-LHC, 
we take them to be constant over the entire range.
For comparison, we also show results for the same model points, but with more conservative efficiencies, in \figsref{scenario_1_charm_fraction_MT2_1500_Eff_c_30}{scenario_1_charm_fraction_MT2_1500_Eff_c_30_Rc}.

\subsection{Discussion} \label{sec:discussion}

The charm fractions displayed
in~Figs.~\ref{fig:scenario_1_charm_fraction_MT2_1500_Eff_c_50}--\ref{fig:scenario_1_charm_fraction_MT2_1500_Eff_c_30_Rc}
contain both SM and supersymmetric contributions.
As noted in \secref{tagging}, our analysis overestimates the number of charm jets from gluon splitting compared to realistic detectors,
since with \texttt{Delphes}, a jet containing a $c\bar{c}$ pair is labelled as a charm jet.
This occurs in both supersymmetry and SM events, but since gluon splitting is more important in the SM, 
the effect on the SM charm fraction is more pronounced.

The charm fraction of supersymmetric models can in principle be lower than in SM events. 
This requires a low gluino mass, which is not of much interest to us, since discovery in this case occurs based on the total number of events.
In the pure supersymmetry sample, the charm fractions of $\nsq=8$ and $\nsq=4$ are identical (assuming the other three parameters are equal).
Note that, while bino-mediated $t$-channel processes give an $\Order(1\%)$ modification of the cross section
and are thus negligible, processes involving winos may give an $\Order(10\%)$ contribution even if the winos are as heavy as the squarks.
Therefore, in the presence of left-handed squarks, our results assume an extremely large wino mass.
Still, our analysis can be straightforwardly generalized to include winos, with little qualitative changes. 
Charm squarks would mainly decay to strange quarks in this case and vice versa.

As mentioned above, estimating the number of charm jets from SM processes is nontrivial.
Beyond this theoretical difficulty, in standard experimental analyses, collinear charm pairs from gluon splitting
are typically merged into a single jet, but jets containing two heavy quarks are subsequently discarded~\cite{Ilten:2017rbd}.
New approaches for heavy flavor tagging were proposed recently to address this problem~\cite{Ilten:2017rbd}.
The intrinsic charm fraction of the proton is another potential source of charm quarks that is hard to estimate~\cite{Brodsky:2015fna}.

Fortunately, these theoretical uncertainties can be straightforwardly circumvented by measuring the charm content of the SM background in the data.
For the Meff-2j-3100 SR analysis, the background is dominated by $Z$ + jets, 
and the charm fraction can be measured in the analogous sample with the $Z$ decaying leptonically.
In fact, $Z+c$ production with leptonic $Z$ decays has been used by CMS for training some of their charm taggers~\cite{CMS-PAS-SMP-15-009}.
Thus, one can extract the numbers of both charm and non-charm jets in the sample of invisible-$Z$ decays,
and by subtracting them, obtain the purely supersymmetric charm fraction.
While  this will be subject to larger experimental uncertainties, the theory systematics will be significantly improved.


\section{Conclusions} \label{sec:concl}

The exclusion limits on superpartner masses from ATLAS and CMS are fast approaching the discovery reach of the LHC.
The fraction of charm quarks in jet plus missing energy events provides a new handle on superpartner production,
and may increase the sensitivity of LHC searches to squark-pair production.
While we have only studied here squark pair production,
the charm fraction in gluino-pair production is of interest too as this process is flavor democratic for degenerate squarks.

We did not address here the production of top and bottom squarks.
Because of their relatively large Yukawa couplings, these are likely to be split in mass from the first- and second-generation squarks.
Furthermore, because the bottom and top content of the proton is negligible,
their production is mainly gluon-mediated, and to a good approximation, independent of the gluino mass.
Thus, while their discovery would yield additional information,
it is orthogonal to our discussion here.
We also neglect winos and higgsinos in this study.
The latter have little effect on first- and second-generation squark production.
Winos, on the other hand, mediate $t$-channel squark production,
and would alter our results unless they are very heavy.

We have argued that the charm fraction can be used to disentangle different model points with similar kinematics.
We note that, while event kinematics are largely governed by the squark and bino mass,
they have some sensitivity to the gluino mass as well,
since how central or forward the events are depends on the weight of $t$-channel gluino processes.
This suggests that measurements of the charm fraction may be optimized by
a judicious choice of kinematic cuts in order to extract the gluino mass.

Here we studied models with mass-degenerate up and charm squarks.
While plausible, this is by no means mandatory.
With mass splittings between the squarks, the fermion Cabibbo mixing 
will typically translate into up-charm mixing of the left-handed squarks;
for concrete spectra, see, e.g., Ref.~\cite{Ierushalmi:2016axs}.
Measuring the charm fraction will yield information on the squark flavor composition.


\section*{Acknowledgments}

We thank Eilam Gross, Heather Gray, Roni Harnik, Michelangelo Mangano, Gilad Perez, Yoram Rozen, Jonathan Shlomi, and Shlomit Tarem for discussions.
We also thank David Cohen for computing support.
This work was performed in part at the Aspen Center for Physics,
which is supported by National Science Foundation grant PHY-1607611,
and Y.S.~was partially supported by a grant from the Simons Foundation.
S.I. and G.L. acknowledge the hospitality of the Mainz Institute for Theoretical Physics, where part of this work was completed.
S.I. is supported in part at the Technion by a fellowship from the Lady Davis Foundation.
Research supported by the Israel Science Foundation (Grant No.~720/15),
by the United-States-Israel Binational Science Foundation (BSF) (Grant No.~2014397),
and by the ICORE Program of the Israel Planning and Budgeting Committee (Grant No.~1937/12).

\bibliography{charm.bib}

\end{document}

%% file: feyn_bkg_zj_gluon.tex
		\begin{fmffile}{diagram_Z_1_Decay_gluonSplitting}
			\begin{fmfgraph*}(150,100)
				
				\fmfleft{i1,i2}
				\fmfright{o1,o2,o3,o4}
				\fmf{fermion,tension=2}{i1,v1}
				\fmf{fermion,tension=2}{v1,i2}
				\fmf{gluon,tension=2}{v1,v2}
        \fmf{fermion}{o4,v2}
				\fmf{phantom}{v2,o1}
				\fmffreeze
				\fmf{phantom}{v2,v3}
				\fmf{phantom}{v2,v3,o1}
				\fmffreeze
        \fmf{fermion}{v2,o1}
				\fmf{boson,label=$Z$,label.side=left}{v3,v4}
				\fmf{fermion,label=$\nu$,label.side=right,label.dist=5}{v4,o2}
				\fmf{fermion,label=$\bar{\nu}$,label.side=left}{o3,v4}
				\fmfdotn{v}{4}
			\end{fmfgraph*}
		\end{fmffile}

%% file: feyn_bkg_zj_quark.tex
		\begin{fmffile}{diagram_Z_1_Decay_ProtonQuark}
			\begin{fmfgraph*}(150,100)
				\fmfleft{i1,i2}
				\fmfright{o1,o2,o3,o4}
				\fmf{gluon,tension=2}{i1,v1}
				\fmf{fermion,tension=2}{i2,v1}
				\fmf{fermion,tension=2}{v1,v2}
				\fmf{gluon}{v2,o4}
				\fmf{phantom}{v2,o1}
				\fmffreeze
				\fmf{phantom}{v2,v3}
				\fmf{phantom}{v2,v3,o1}
				\fmffreeze
				\fmf{fermion}{v2,o1}
				\fmf{boson,label=$Z$,label.side=left}{v3,v4}
				\fmf{fermion,label=$\nu$,label.side=right,label.dist=5}{v4,o2}
				\fmf{fermion,label=$\bar{\nu}$,label.side=left}{o3,v4}
				\fmfdotn{v}{4}
			\end{fmfgraph*}
		\end{fmffile}

%% file: feyn_bkg_wj_tau.tex
		\begin{fmffile}{diagram_W_1_tauDecay}
			\begin{fmfgraph*}(150,100)
				
				\fmfleft{i1,i2}
				\fmfright{o1,o2,o3}
				\fmf{gluon,tension=2}{i1,v1}
				\fmf{fermion,tension=2}{i2,v1}
				\fmf{fermion,tension=2}{v1,v2}
				\fmf{fermion}{v2,o3}
				\fmf{phantom}{v2,o1}
				\fmffreeze
				\fmf{phantom}{v2,v3}
				\fmf{phantom}{v2,v3,o1}
				\fmffreeze
				\fmf{photon,label=$W$,label.side=left}{v2,v3}
				\fmf{fermion,label=$\tau_{\mathrm{had}}$,label.side=right,label.dist=5}{v3,o2}
				\fmf{fermion,label=$\bar{\nu}$,label.side=left}{o1,v3}
				
				\fmfdotn{v}{3}
			\end{fmfgraph*}
		\end{fmffile}

%% file: feyn_bkg_wj_lostlepton.tex
		\begin{fmffile}{diagram_W_1_lostlepton}
			\begin{fmfgraph*}(150,100)
				\fmfleft{i1,i2}
				\fmfright{o1,o2,o3,o4}
				\fmf{gluon,tension=2}{i1,v1}
				\fmf{fermion,tension=2}{i2,v1}
				\fmf{fermion,tension=2}{v1,v2}
				\fmf{gluon}{v2,o4}
				\fmf{phantom}{v2,o1}
				\fmffreeze
				\fmf{phantom}{v2,v3}
				\fmf{phantom}{v2,v3,o1}
				\fmffreeze
				\fmf{fermion}{v2,o1}
				\fmf{boson,label=$W$,label.side=left}{v3,v4}
				\fmf{fermion,label=$\cancel{\ell}$,label.side=right,label.dist=5}{v4,o2}
				\fmf{fermion,label=$\bar{\nu}_{\ell}$,label.side=left}{o3,v4}
				\fmfdotn{v}{4}
			\end{fmfgraph*}
		\end{fmffile}

%% file: feyn_democratic_production.tex
	$\begin{array}{ccc}
	\begin{fmffile}{diagram_Q_3}     
	\begin{fmfgraph*}(120,90)
	\fmfleftn{i}{2}
	\fmfrightn{o}{2}
	\fmfdotn{v}{2} 
  \fmf{fermion}{v1,i1}  
    \fmf{fermion}{i2,v1}
	\fmf{gluon}{v1,v2}
	\fmf{scalar, label=$\tilde{q}^{*}$}{o1,v2} 
	\fmf{scalar, label=$\tilde{q}$}{v2,o2} 
	\end{fmfgraph*} 
	\end{fmffile} 
	&\begin{fmffile}{diagram_ggss}     
	\begin{fmfgraph*}(120,90)
	\fmfleftn{i}{2}
	\fmfrightn{o}{2}
	\fmfdotn{v}{1} 
	\fmf{gluon}{v1,i1}  
	\fmf{gluon}{i2,v1}
	\fmf{scalar, label=$\tilde{q}^{*}$}{o1,v1} 
	\fmf{scalar, label=$\tilde{q}$}{v1,o2} 
	\end{fmfgraph*} 
	\end{fmffile} 
	& \begin{fmffile}{diagram_tChannel_1}     
	\begin{fmfgraph*}(120,90)
	\fmfleftn{i}{2}
	\fmfrightn{o}{2}
	\fmfdotn{v}{2} 
	\fmf{gluon}{i1,v1}  
	\fmf{gluon}{i2,v2}
	\fmf{scalar, label=$\tilde{q}^{*}$}{o1,v1} 
	\fmf{scalar, label=$\tilde{q}$}{v2,o2} 
	\fmf{scalar, label=$\tilde{q}$}{v1,v2}
	\end{fmfgraph*} 
	\end{fmffile}
	\end{array} $

%% file: feyn_undemocratic_production.tex
%
%
%
	
	$\begin{array}{cccc}
	\begin{fmffile}{diagram_Q_1_Decay}
	\begin{fmfgraph*}(120,90)
	\fmfleftn{i}{2}
	\fmfrightn{o}{2}
	\fmfdotn{v}{2}
	\fmf{fermion, label=$q^{\prime}$}{i1,v1}
	\fmf{fermion, label=$q$}{i2,v2}
	\fmf{gluino, label=$\tilde{g}$,label.dist=10}{v2,v1}
	
	\fmf{scalar, label=$\tilde{q}$}{v2,o2}
	
	\fmf{scalar, label=$\tilde{q}^{\prime}$}{v1,o1}
	
	\end{fmfgraph*}
	\end{fmffile}
	\end{array} $